\documentclass[preprint]{revtex4-1}
\usepackage{graphicx}

\usepackage{natbib}

\usepackage{graphics}

\usepackage{natbib}

\newcommand{\be}{\begin{equation}}
\newcommand{\ee}{\end{equation}}
\newcommand{\nn}{\mbox{} \nonumber \\ \mbox{} }
\newcommand{\ba}{\begin{eqnarray}}
\newcommand{\ea}{\end{eqnarray}}

\newcommand{\E}{{\bf E}}
\newcommand{\B}{{\bf B}}

\renewcommand{\v}{{\bf v}}

\newcommand\eg{{e.g.,\ }}
\newcommand\cf{{cf.\ }}

\newcommand{\Bf}{{magnetic field}}
\newcommand{\Bfs}{{magnetic fields}}
\newcommand{\Ef}{{electric  field}}
\newcommand{\Efs}{{electric fields}}

\newcommand{\NS}{neutron star}
\newcommand{\NSs}{{neutron stars}}
\newcommand{\EM}{electromagnetic}
\newcommand{\BH}{{black hole}}
\newcommand{\BHs}{{black holes}}
\newcommand{\Sc}{Schwarzschild}
\newcommand{\ms}{magnetosphere}
\newcommand{\mss}{magnetospheres}
\newcommand{\Fermi}{{\it Fermi}}

\begin{document}

\title{Schwarzschild  black holes as unipolar inductors: expected \EM\ power of a merger }

\author{Maxim Lyutikov\\
Department of Physics, Purdue University, \\
 525 Northwestern Avenue,
West Lafayette, IN
47907-2036 }

\begin{abstract}
The motion of a  Schwarzschild  black hole with velocity $v_0 = \beta_0 c$   through a  constant \Bf\  $B_0$  in vacuum induces  a component of the  \Ef\  along   the  magnetic  field,  generating a non-zero  second  Poincare  electromagnetic  invariant $ ^* F \cdot F \neq 0$.  This will  produce (\eg via radiative effects and  vacuum breakdown) an  electric charge density  of the order of $\rho _{\rm ind}= B_0 \beta_0 /( 2  \pi e R_G)$, where  $R_G =   2 G M/c^2$ is the Schwarzschild radius and $M$ is the mass of the \BH;  the  charge density $\rho _{\rm ind}$ is similar to the Goldreich-Julian density.

 The magnetospheres of  moving \BHs\ resemble in many respects  the \mss\ of rotationally-powered pulsars, with  pair formation fronts and  outer gaps, where the sign of the  induced charge changes. As a result, 
the black hole  will generate bipolar electromagnetic jets each consisting of  two counter-aligned current flows (four current flows  total), each carrying an 
  electric current  of the order $I \approx e   B_0 R_G  \beta_0$. The electromagnetic power of the jets is $L \approx (G M)^2 B_0^2 \beta_0^2/c^3$;  for a particular case of  merging black holes the  resulting Poynting power is  $ L \approx {(G M)^3 B_0^2 /( c^5 R) }$,  where  $R$ is the radius of the orbit. 

   In addition, in  limited regions near the horizon the first electromagnetic invariant  changes sign, so that the induced \Ef\ becomes larger than the  \Bf, $E>B$. As a result, there will be  local dissipation of the \Bf\ close to the horizon, within a region with the radial extent
    $\Delta R \approx R_G \beta_0$. 

The total energy loss from a system of merging \BHs\ is a sum of two components with similar powers, one due to the rotation of space-time within the orbit, driven by the non-zero angular momentum in the system, and the other due to the  linear motion of the \BHs\ through the  \Bf.

Since the resulting electrodynamics is in many respects similar to pulsars, 
 merging black holes may generate coherent radio and high energy 
  emission beamed  approximately along the orbital normal. In addition,  merging \BHs\ may produce  observable wind-driven cavities.  

 \end{abstract}

\maketitle

 
\section{Introduction}
Observations of an  \EM\ signal  accompanying \BH\ merger is most desirable, as it will provide  crucial  information on the location and the physical properties of the event. Two types of \EM\  signal can be expected from merging \BHs. First, merging \BHs\ induce perturbations in the surrounding gas \cite{2002ApJ...567L...9A,2010ApJ...709..774K,2005ApJ...622L..93M,2010ApJ...711L..89V}. The resulting \EM\ signal is  then subject to great uncertainty and   naturally  depends on the complicated non-linear fluid behavior of the system.  One of the problems  is that at late stages of the merger there should be little gas inside the orbit  since the  timescale for shrinkage of the binary orbit by gravitational wave radiation becomes shorter than the timescale for mass inflow due to viscose stresses in the disk \cite{2005ApJ...622L..93M}. It is then hard to excite transient dissipative processes in a faraway accretion disk.

 Alternatively, external gas can support electric currents that create large scales \Bfs. Motion of \BHs\ in this externally supplied \Bf\ can then lead to electromagnetic extraction of energy. Qualitatively,  {\it there are two distinct possibilities for the  electromagnetic extraction of energy from spiraling \BHs.} First, the system of two orbiting  \BHs\ 
possesses non-zero angular momentum, which induces rotation of space time.  Rotating space-time can generate \EM\ outflows, in a manner similar to the classical Faraday disk.
This is the physics behind the  Blandford \& Znajek
 \cite{Blandford:1977} process of extracting  the rotational power of a \BH. Below we refer to this mechanism as the   Faraday disk mechanism.  This mechanism has been previously considered in Ref. \cite{2010arXiv1010.6254L}.  For a given \Bf, the  resulting power can be estimated using the Faraday disk scaling,
 $L _{EM,F}\sim  R^2 B_0^2 c (\Omega R/c)^2$, where  $R$ is the orbital radius and $\Omega$ is   the typical angular velocity of the rotation of the space-time within the {\BHs}' orbit,
 $\Omega \sim  { (GM)^{3/2} /( c^2 R^{5/2}})$. The resulting  Poynting flux is then
\be 
L_{EM,F} \approx {  G^3 M^3 \over c^5 R} B_0^2
\label{L}
\ee
 
 In a separate, physically distinct process, which we consider in this paper, the linear motion of  a \Sc\ \BH\ will generate the electric potential drop across the \BH, so that a \BH\ will effectively  operate as a unipolar inductor,  in a way somewhat similar to the planet Io moving in Jupiter's \Bf\  \cite{1969ApJ...156...59G}.  Classically, 
the motion of a conductor though the  \Bf\ generates in the frame of a conductor an \Ef\ $\E = - \v \times \B$. This induced \Ef\ will generally have a normal component to the surface of the conductor. As a result, surface charges will be generated; they will produce their own \Ef, now with a  component parallel to the initial \Bf.  This \Ef\ drives currents along \Bf\ lines; dissipation of these currents is responsible for non-thermal radio through X-ray emission of the Jupiter \ms.
 An important  difference of the unipolar induction mechanism considered in the present paper  from the classical unipolar inductor is that in the case of a \BH\ {\it no physical charges are needed to produce $E_\parallel$}. Parallel \Ef\ appears in complete vacuum due to the curvature of space.

A qualitative estimate of the resulting Poynting power may be obtained from the following reasoning. 
For a conductor of length $l$ moving with velocity $v_0 = \beta_0 c$  through \Bf\ $B_0$ the resulting potential difference $\Delta \Phi \sim \beta_0  L B$.  If the resulting outflow is relativistic, 
 the \EM\ power can be estimated as $L _{EM} \sim  \Delta \Phi^2  c = \beta_0 ^2 B_0^2 l^2 c$  (see also \cite{2010arXiv1012.2872M}). In the case of orbiting  \BH, estimating $l \sim R_G$ (the Schwarzschild radius) and $ \beta_0 = \sqrt{R_G /R}$, the expected power of unipolar inductor turns out to be the same as that of the Faraday disk mechanism, Eq. (\ref{L}). 
 
Thus, the estimates of the \EM\ powers due to rotation of the space-time and due to linear motion of a \BH\ with Keplerian velocity turn out to be similar, given by Eq. (\ref{L}); we  view this as a coincidence. Though in both cases the power eventually comes from the inductive \Ef,   the underlying physics is different in the two case. One mechanism requires non-zero angular momentum, while  the other does not. A total energy loss from a system of merging \BHs\ is a sum of two components with similar powers, one due to the rotation of space-time within the orbit, another due to linear motion of the \BHs\ through \Bf.
 
  Previously,   in Refs. \cite{2010PhRvD..82d4045P,2010Sci...329..927P,2010arXiv1012.5661N} a number of  force-free simulations  of \BH\ \mss\ were performed. In the case of the orbiting \BHs\ the  authors mostly studied the \EM\ power due to linear motion of \BH,  and not due to rotation of the space-time within the orbit. The present paper offers explanations and parameter scalings  of these numerical simulation.

\section{Static \EM\ fields in \Sc\ metric }

Consider vacuum stationary homogeneous \EM\ fields in   \Sc\ metric.
  Though the structure of  a constant \EM\ fields in \Sc\  metric is well known  \citep{Wald,Hanni1,Ernst,Beskin}, here  we briefly re-derive it here for completeness.
Adopting a  \Sc\ metric 
 \be
 ds^2 = - \alpha^2 dt^2 + {1\over  \alpha^2} dr^2 + r^2 \left( d\theta^2 +  \sin ^2 \theta d\phi^2\right)
 \ee
 where $\alpha = \sqrt{1- 2M/r}$, the relevant vacuum  Maxwell equations 
 \be
 \partial_\nu  \left( \sqrt{-g} F^{\mu \nu} \right) =0, 
 \label{Maxwell}
 \ee
 ($F_{\mu \nu} = A_{[\mu, \nu]}$ is Maxwell tensor) for the non-vanishing components of the vector potential $A_0 (r, \theta')$ and $A_\phi (r, \theta)$ 
 (here $\theta' $ and $\theta$ are angles with respect to the axes aligned with the electric and \Bf\ at infinity) give
 \ba &&
  \alpha^2 \partial_r ( r^2 \partial_r A_0) + {1\over \sin \theta'}  \partial_{\theta' } ( \sin \theta' \partial_{\theta'} A_0)=0
 \nn &&
 r^2 \partial_r \alpha^2 \partial_r A_\phi +\sin \theta \partial_\theta \left( {1 \over \sin \theta} \partial _\theta A_\phi  \right) =0.
 \label{GS}
\ea
(In vacuum, the same relations hold for the dual Maxwell tensor $^* F_{\mu \nu}$; in that case the equations for $A_0$ and $A_\phi$ switch.)
The potentials  corresponding to constant fields at infinity are
\ba && 
A_0 = 
 E_0 (r - 2M) \cos \theta'    
\nn && 
A_\phi = {B_0 \over 2} r^2 \sin^2 \theta
\ea
The \EM\ fields 
\ba && 
\E = (1/ \alpha) \nabla A_0 = E_0 \left( { \cos \theta' }  {\bf e}_r  -  \alpha  \sin \theta'  {\bf e}_\theta \right)
\nn &&
\B =  {\nabla  \times A_\phi   {\bf e}_\phi} = B_0 \left(  { \cos \theta }   {\bf e}_r  -   \alpha  \sin \theta  {\bf e}_\theta \right)
\label{BB}
\ea
where  $\nabla$ is a covariant derivative, with a corresponding  unit radial vector 
$
{\bf \hat{e}}_r = \alpha \partial_r
$.
The structure of the electric and \Bfs\ is the same, as follows from the duality transformation in vacuum.

The above relations can be obtained in a more conventional way by using the  alternative $3+1$ formulation  of the General Relativity   \cite{ThornMembrane}, in which case  the  Maxwell equations in the more general Kerr metric take  the form
 \ba && 
 \nabla \cdot \E =  4 \pi \rho
 \nn &&
 \nabla \cdot \B=0
 \nn &&
 \nabla  \times ( \alpha \B) = 4 \pi \alpha {\bf j} + D_t \E
 \nn &&
 \nabla   \times ( \alpha \E) = - D_t \B
 \label{maxw1}
 \ea
  where $D_t = \partial _t - {\cal L} _{\vec{\beta}} $ is the total time derivative, including Lie derivative along the velocity of the zero angular momentum observers (ZAMOs).
For \Sc\ metric $\vec{\beta}=0$ and in the stationary case  $D_t \equiv 0$. 
 Electromagnetic fields in Eq. (\ref{maxw1}) are those measured by a stationary local observer in terms of local time. The fields $\alpha \E$ and $\alpha \B$ are those measured by a local observer in terms of \Sc\ time $t$.
 
 %Eqs. (\ref{maxw}) are written in the $3+1$ splitting formulation of general relativity \cite{ThornMembrane}.
%Electromagnetic fields in Eq. (\ref{maxw}) are those measured by a stationary local observer in terms of \Sc\ coordinates.
%Also recall that   \Sc\ metric  is expressed in terms of quantities measured by an observer at infinity.
Since we expect that the  non-zero charge will eventually be created, we give here the complete Laplace equation for the electric potential $A_0$ in presence of  non-zero charge density 
  \ba &&
  \nabla {1\over \alpha} \nabla A_0= 4\pi \rho
  \nn &&
{1 \over r^2}  \partial_r (r^2  \partial_r A_0) + {1\over  \alpha^2  r \sin \theta'} \partial _{\theta} (  \sin \theta \partial _\theta A_0 ) + {1\over   \alpha^2 r^2 \sin^2 \theta} \partial _\phi^2 A_0 = 4 \pi {  \rho \over \alpha}
\label{rho}
\ea
\cf \cite[][Eq. 10]{Linet}.

The shape of the field lines  (\ref{BB}) is given by
\be
\sin \theta = 2  e^{ -2 (1-\alpha)} { 1-\alpha \over 1+ \alpha } {r_{\perp,0} \over M}
\label{shape}
\ee
where $r_{\perp,0} $ is the radial cylindrical distance from the axis of a given field line at infinity.  The last  field line that intersects the \BH\ at $\theta =\pi/2$ has initial $r_{\perp,0}  = e^2 M /2$ and is given by  $\sin \theta= 2 { e^{  2 \alpha } \over (1+\alpha)} {  M \over r}$.
Note, that though the surfaces of constant magnetic flux are cylinders $r \sin \theta={\rm constant}$ (corresponding to $A_\phi =$ constant),  on the horizon the \Bf\  in \Sc\ coordinates becomes radial. This is naturally impossible at the point $ r=2M, \theta = \pi/2$, but at that  point the  \Bf\  is zero. This seeming inconsistency can be resolved in terms of embedding diagrams \cite{Hanni1}.

By construction  fields (\ref{BB}) correspond to $\nabla \cdot \E=0$, so  any local observer  would measures zero charge  density.
 On the other hand, judging by the shape of \Ef\ lines  in coordinates $r-\theta$, the observer at infinity will infer a dipolar-like charge distribution
\be
\rho_{eff} = \nabla_{\rm nc} \cdot \E= 2 \left(1 - \alpha \right) {\cos \theta' \over r}  E_0
\ee
Here $\nabla_{\rm nc}$ denotes the non-covariant differentiation with respect to coordinates $r-\theta$.
The effective charge is concentrated near the \BH\, so at large distances the \BH\ effectively has a surface charge
\be
\sigma_{eff}  = {1\over 4 \pi} \int_{2M}^\infty  \rho_{eff} dr = {\ln (e/2) \over \pi} E_0 \cos \theta'
\ee
We stress that there are no physical charges present: the shape of \Ef\ lines in the chosen metric is  modified  by  gravity, not electric charges. Also, the shape of field lines and the value of the effective surface density depends on the choice of coordinates and a given set of observers.

  \begin{figure}[h!]
   \includegraphics[width=0.99\linewidth]{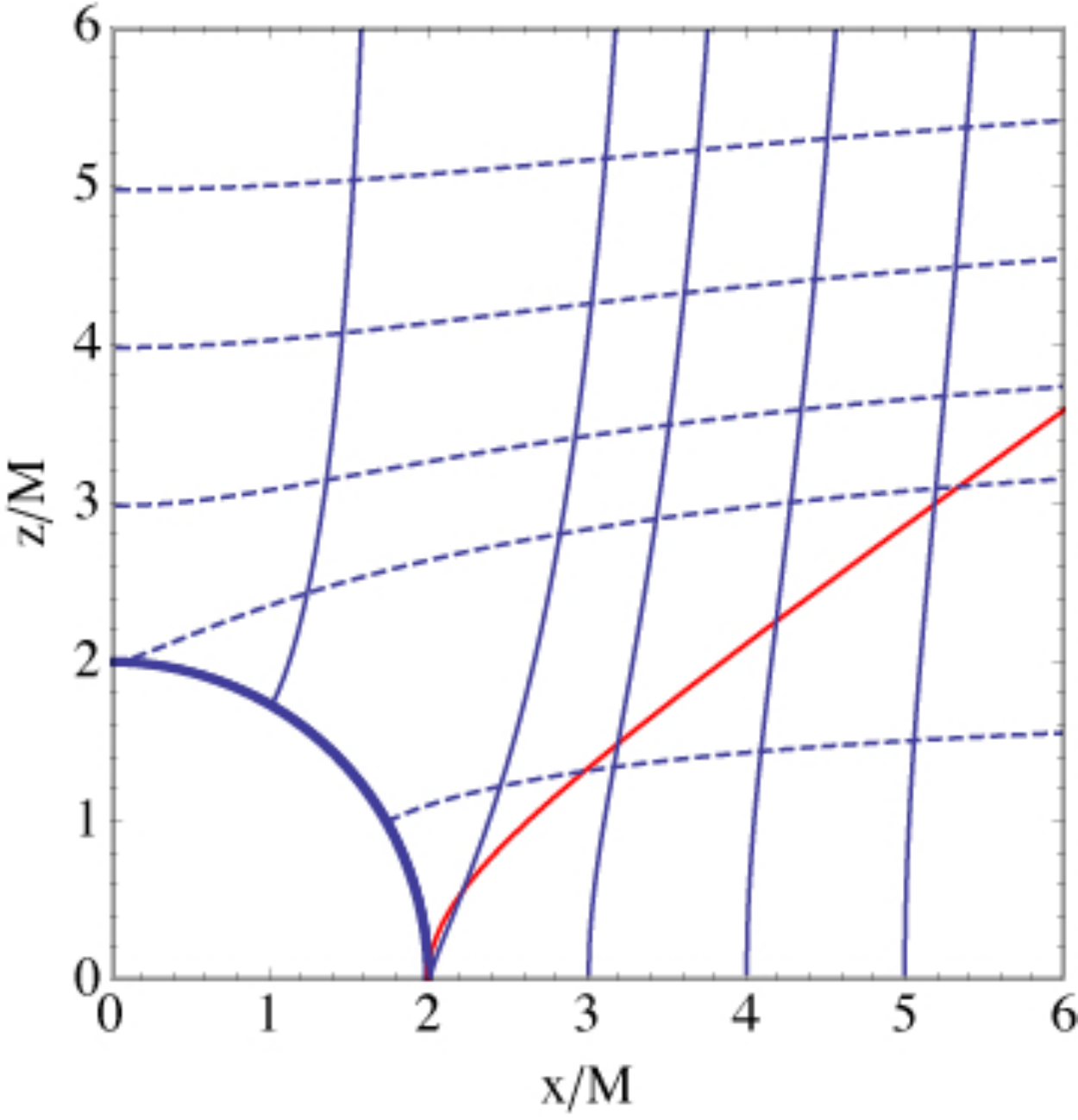}
   \caption{Shape of  magnetic   (solid lines aligning  with $z$ axis) and electric  (dashed lines, aligning with $x$ axis) field  lines in the $x-z$ plane for vacuum \Sc\ \BH.  Red curve (in the online color image) starting at the point $\theta = \pi/2$, $r=2M$ is the shape of the outer gap, where the sign of  the induced charges changes.  }
 \label{mainpicture}
 \end{figure}

 \begin{figure}[h!]
   \includegraphics[width=0.9\linewidth]{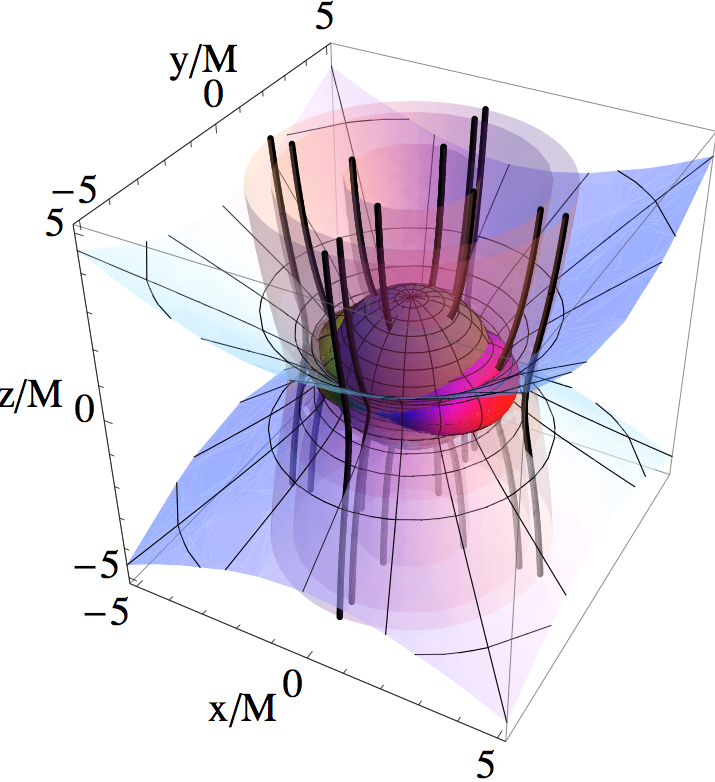}
   \caption{ 3D view of the \BH\ magnetosphere. Magnetic field (thick solid lines together with magnetic surfaces)  is along $z$ axis at large distance, the \BH\ is moving along the  $y$ direction. The central sphere of radius $r=2 M$ is the \Sc\ radius. Two limited regions near the equatorial plane approximately aligned with the $x$  axis are bounded by the surfaces $E=B$. On the nearly conical surfaces originating at the  magnetic equator the induced charge density is zero. This particular plot is for $\beta_0= 1/2$.   }
 \label{bigpicture}
 \end{figure}

\section{\BH\ in \EM\ fields orthogonal at infinity. }

Let us next assume that in a particular  reference frame at large distances from the \BH\ the  \Bf\ is along $ -  {\bf e}_z$ direction, while in this reference frame the  \Ef\ is zero.  A \BH\ is moving orthogonally to   \Bf\ with velocity $\beta_0 = v_0 / c $ along $y$ direction. In the frame of the \BH,  at large distances $ r \gg M$,  there is then a  static  \Ef\ $\E_0= - \v \times \B_0 =\beta_0 B_0 {\bf e}_x $, $E_0 = \beta_0 B_0$. (In this section  we define $B_0$ as the value of \Bf\ in the frame where \BH\ is at rest; it is related to the value in the frame where \Ef\ is vanishing by a simple Lorentz transformation). 
Thus, in the frame of the hole  the  electric and \Bfs\ are given by Eq. (\ref{BB}), where $\theta' $ is a polar angle with respect to the $x$ axis,
  $ \cos \theta'  = \sin \theta \cos \phi$.  
  The \EM\ invariants are 
\ba && 
B^2-E^2 = B_0^2 (1-\beta_0^2) + 2  \left(\beta_0^2 ( \cos^2 \theta \cos^2 \phi +\sin^2 \phi) - \sin ^2 \theta  \right) {  B_0^2 M \over r}
\nn &&
\E \cdot  \B \equiv -  { {\rm Det } |F_{\mu\nu} |\over \sqrt{-g}} = -  \cos \phi \sin 2 \theta \beta_0 B_0^2{  M\over r}
\label{inv}
\ea
while the parallel \Ef\ is 
\be
E_\parallel =  -  \cos \phi \sin 2 \theta   {  M\over r \sqrt{1-2 \sin ^2 \theta M /r} } E_0
\ee

Eqns. (\ref{inv}) highlight two important points. The second Poincare  \EM\ invariants is generally non-zero, $\E \cdot \B \not \equiv 0$. The first  Poincare  invariant changes sign  on the surface
\be
r= { 2 M \left( \sin ^2 \theta - \beta_0^2 ( \cos ^2   \theta \cos ^2 \phi - \sin ^2 \phi ) \right)\over 1 -\beta_0^2}
\label{2}
\ee
 This occurs  close to the horizon around  points  $\{\theta \approx \pi/2, \phi \approx 0, \pi\}$ within a region ${  \Delta r \over 2 M}  \approx \beta_0^2/(1-\beta_0^2)$, see Fig. \ref{bigpicture}.

\begin{figure}[h!]
   \includegraphics[width=0.49\linewidth]{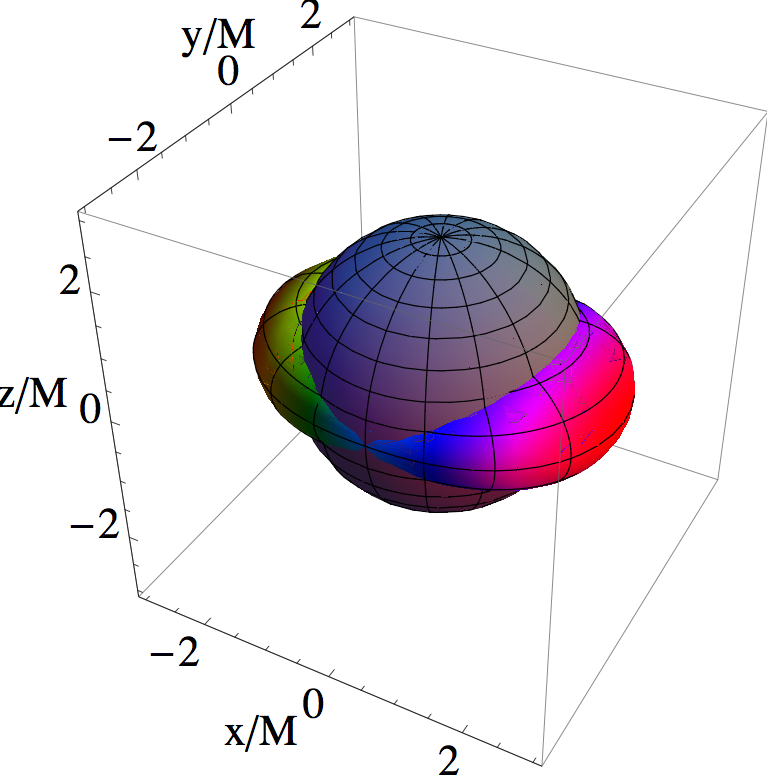}
 \includegraphics[width=0.49\linewidth]{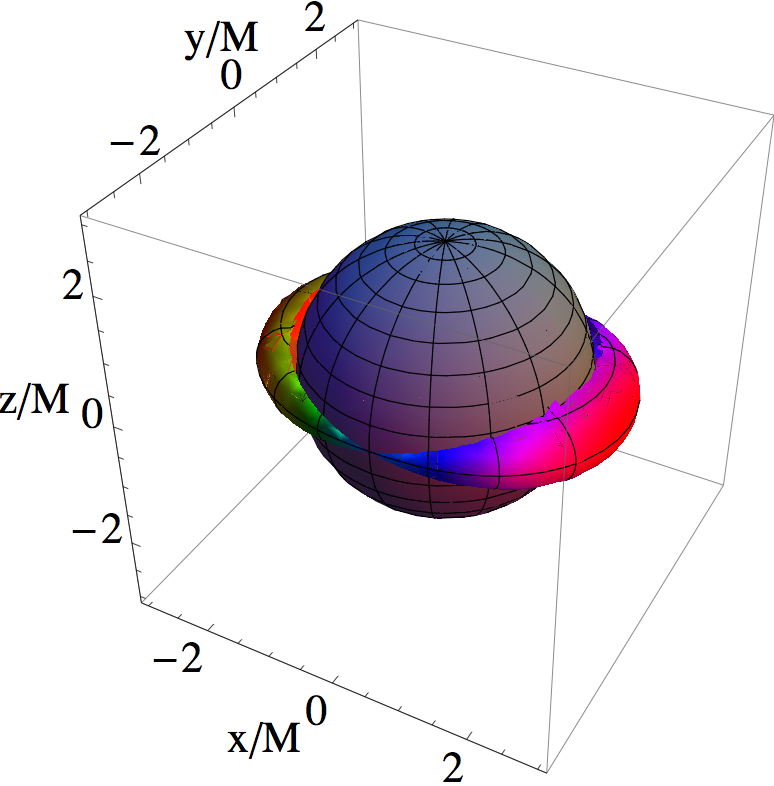}
   \caption{3D view of the surfaces bounding the regions where $E>B$ ({\it Left Panel}) and  $E_\perp>B$ ({\it Right Panel})  for $ \beta_0 =0.5$.   The central sphere of radius $r=2 M$ is the \Sc\ radius. 
   False colors (in the online version) in this  figure and Figs.  4 and 9 are chosen for presentation purpose only.  }
 \label{EeqB}
 \end{figure}

We expect that in astrophysical environment  $E_\parallel$ will be decreased by pair creation (see \S \ref{rr}). Still, the perpendicular component of the \Ef\   $E_\perp$ becomes larger than the \Bf\ at points
\be
\left( 1- { 2 M \sin ^2 \theta \over r } \right)^2 = \beta_0^2 \alpha \left(1  - { 2 M \sin ^2 \theta  \sin ^2 \phi \over r } \right) 
\label{1}
\ee
The region bounded by this  surface  has a shape similar to the one where $E=B$, Eq.  (\ref{2}), see Fig. \ref{EeqB}.

Since we expect that a plasma will be generated due to vacuum breakdown, we can define the plasma drift velocity (assuming  $\beta_d<1$, see below)
 \be
{\vec{ \beta_d}} = \E \times \B /B^2 =  {\alpha  \over 1- 2 M \sin ^2 \theta/r}  \left(  
{\alpha} \sin \theta \sin \phi {\bf e}_r +
 \cos \theta   \sin \phi {\bf e}_\theta +
 \cos \phi  {\bf e}_\phi
 \right) \beta_0 
\ee
It is directed along $y$ axis at large distances, but becomes complicated near the hole.  Most importantly, the value of $v$ is not bound. Close to the points 
$\theta =\pi/2, \phi = 0, \pi$,  the velocity diverges  as $\beta_d \approx \beta_0 /\alpha$. This is related to the previously mentioned fact that the second \EM\ invariant changes sign close to these points.

\section{Magnetospheres of moving  \Sc\ {\BH}s}
 \label{rr}

 \subsection{Induced charge density}
 
 In the previous section we showed that the motion of a \Sc\ \BH\ through \Bf\ in vacuum generates non-zero second Poincare \EM\ invariant (parallel \Ef) and regions where  the first Poincare changes sign, $E>B$.
In vacuum this would have been the end of the story, so to say. In reality, astrophysical plasmas   tolerate  neither large parallel electric field, nor regions of $E>B$: any stray particle will be accelerated by \Ef\ and will produce an electron-positron pair via various radiative effects. The resulting pair plasma will try to screen the initial parallel \Ef, by producing   a charge density required to have  $\E_\parallel=0$. 

\begin{figure}[h!]
   \includegraphics[width=0.49\linewidth]{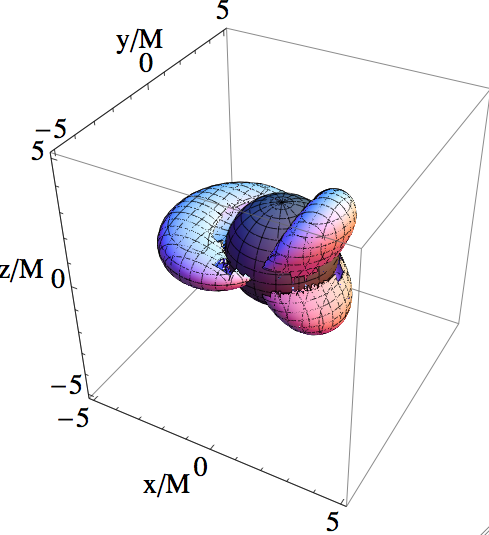}
 \includegraphics[width=0.49\linewidth]{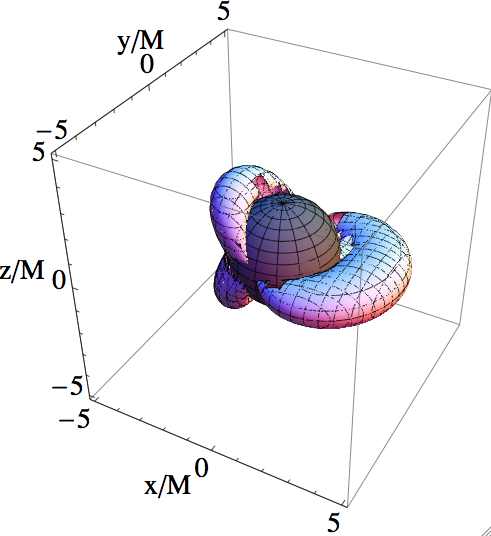}
   \caption{Surfaces of  constant charge density $\rho_{\rm ind} = +|{\rm Const}| $ ({\it Left Panel}) and  $\rho_{\rm ind} = - |{\rm Const}| $  ({\it Right Panel})  for ${\rm Const} =2 \times 10^{-2} \rho_0$  and $ \beta_0 =0.5$.  
   }
 \label{rho-contour}
 \end{figure}

Thus, we expect that around  astrophysical {\BH}s moving in external \Bf\  a non-zero charge density will appear, that satisfies the equation (\ref{rho}) with induced  \Ef\
\be
\E_{\rm ind}= E_\parallel \hat{\bf b}
\label{Epara}
\ee
 where 
 \be
 \hat{\bf b} =  {1 \over \sqrt{1- ( 2 M /r) \sin ^2\theta}} \left( - { \cos \theta }   {\bf e}_r  +   \alpha  \sin \theta  {\bf e}_\theta \right)
 \ee
  is a unit vector along \Bf. 
 % Though the value of the induced charge density $ \rho_{ind} $ can be easily calculated, it's full form is  lengthy and is not given here. 
 The corresponding charge density, see Fig. \ref{rho-contour}
 \be 
 \rho_{\rm ind} = {1\over 4\pi} \nabla \cdot \E={  \alpha \over \pi}   { \left( 1+\left( M/r -3/2\right) \sin ^2 \theta\right) \over (1- ( 2 M /r) \sin ^2\theta)^2}  \cos \phi \sin \theta{M E_0 \over r^2}
 \label{rhoind}
    \ee
The   typical value of charge density $\rho_0$ is 
 \be
 \rho_{0} =  { \beta_0 B_0 \over 4  \pi M } = { B_0  (v_0/ R_G ) \over  2  \pi  c}
 \label{rho0}
 \ee
 This reminds the Goldreich-Julian charge density \citep{GoldreichJulian} if we substitute $v_0/ R_G \rightarrow   \Omega_{eff}$.

At large distance from the  \BH\  the charge density  is 
\be
\rho_{\rm ind} \approx { (1+3 \cos 2 \theta) \cos \phi \sin \theta M^2 \over  r^2} \rho_0
\ee

On the magnetic  equator $\theta = \pi/2$ the charge density is
 \be
 \rho_{ind, \theta=\pi/2} = 2   {  \rho_{0}  \over \alpha}  { M ^2 \over  r^2 }   \cos \phi
 \label{rhoeq2}
 \ee
 It diverges on the horizon. 
  The charge density close to    the horizon $ \rho_{ind,h}$ is 
  \be
 \rho_{ind} =    { \alpha  \rho_{0}  }  {   \cos \phi  \sin  \theta \over \cos^2 \theta}
 \label{rhoh3}
 \ee
  It diverges near the equator.
  
The point $r=2M$, $\theta = \pi/2$ is a special one, as can be seen from the fact that   the limit of Eq. (\ref{rhoh3}) for $\theta \rightarrow \pi/2$ does not coincide with  the limit  $r \rightarrow 2M$ of Eq. (\ref{rhoeq2}). This can be traced to the fact that on the one hand, the lines of force must cross horizon orthogonally \citep[\eg][]{Hanni1}, yet on the other hand they must lay on cylinders    $r \sin \theta={\rm constant}$ (see Eq. (\ref{BB})). These two conditions are inconsistent at the point  $\theta = \pi/2$, $r \rightarrow 2M$.   Recall also that the \Bf\ is zero at this point, Eq. (\ref{BB}), while experiencing a kink, clearly seen  in the embedding diagram  \citep{Hanni1}.

 \subsection{Plasma response}

 Similarly to the case of rotating \NSs, 
 two alternative possibilities exist with  respect to the  dynamical response of the system to the required charge density. First, the overall plasma distribution can provide a {\it static distribution of  the required charge density} and the system will overall be quiet, without strong electromagnetic outflows. In case of pulsars,  this approach is advocated in Ref.  \cite{1979ApJ...227..579M}. This scenario is supported by the fact that direct PIC simulations of the pulsar {\ms}s so far failed to produced an outflow: the system  indeed just relaxes to a nearly steady-state configuration \cite{2001MNRAS.322..209S}. It is expected that secular  instabilities of the resulting charged configurations 
 \citep[\eg diocotron instability,][]{2002A&A...387..520P} may eventually  lead to the formation of the jets (Spitkovsky, priv. comm.).
 
 Alternatively (and this viewpoint is supported by majority of  pulsar theorists), static charge configuration cannot be established on \Bf\ lines connecting to infinity, resulting in the formation of the wind \cite{GoldreichJulian}. Fluid simulation of pulsar \mss\ support this paradigm  \cite{Spitkovsky06}. Since in our case   {\it all} \Bf\ lines are connected to infinity,  large parallel \Efs\  will lead to plasma outflow, which would qualitatively resemble plasma outflow on the open lines of pulsar {\ms}s.  
 
We accept the paradigm that in  the case of \BHs, similar to the pulsar \mss,  parallel \Efs\ will lead to vacuum breakdown, generation of primary beam and  a dense secondary plasma.  In this Section we investigate the resulting \EM\ structure of the \BH\ \mss. 

    \subsection{Pair formation front}

A \BH\  moving with velocity $v_0 = \beta_0 c$ through \Bf\ $B_0$ creates a potential drop of the order $\Delta \Phi \approx \beta_0 r_G B_0$. 
For example, for 
  a \BH\ of  mass $M= 10^6 M_\odot m_6$ moving on the orbit $\xi \gg 1$ times larger than the \Sc\ radius, $R_{\rm orb} = \xi r_G$, in a given \Bf, the resulting potential is
  $\Phi \approx 5 \times 10^{13} \, {\rm V} \left( {B\over 1 {\rm G}} \right)  m_6 \xi ^{-1/2}$. 
  This will result in a particle energy, which is  typically much larger than  the one required to break the vacuum. For example, in pulsar \mss, 
  the pulsar death line corresponds approximately to $\Phi \approx  10^9 $ V \citep[\eg][]{AronsScharlemann,2001ApJ...554..624H}. Though radiative processes and the photon fields around pulsars  and around \BHs\ can be substantially different, in case of \BHs\ the available potential is many order of magnitude larger than the one corresponding to the pulsar death line. This should ensure  the vacuum breakdown and the  formation of the secondary  pair plasma in \BH\ \mss.

\begin{figure}[h!]
   \includegraphics[width=0.99\linewidth]{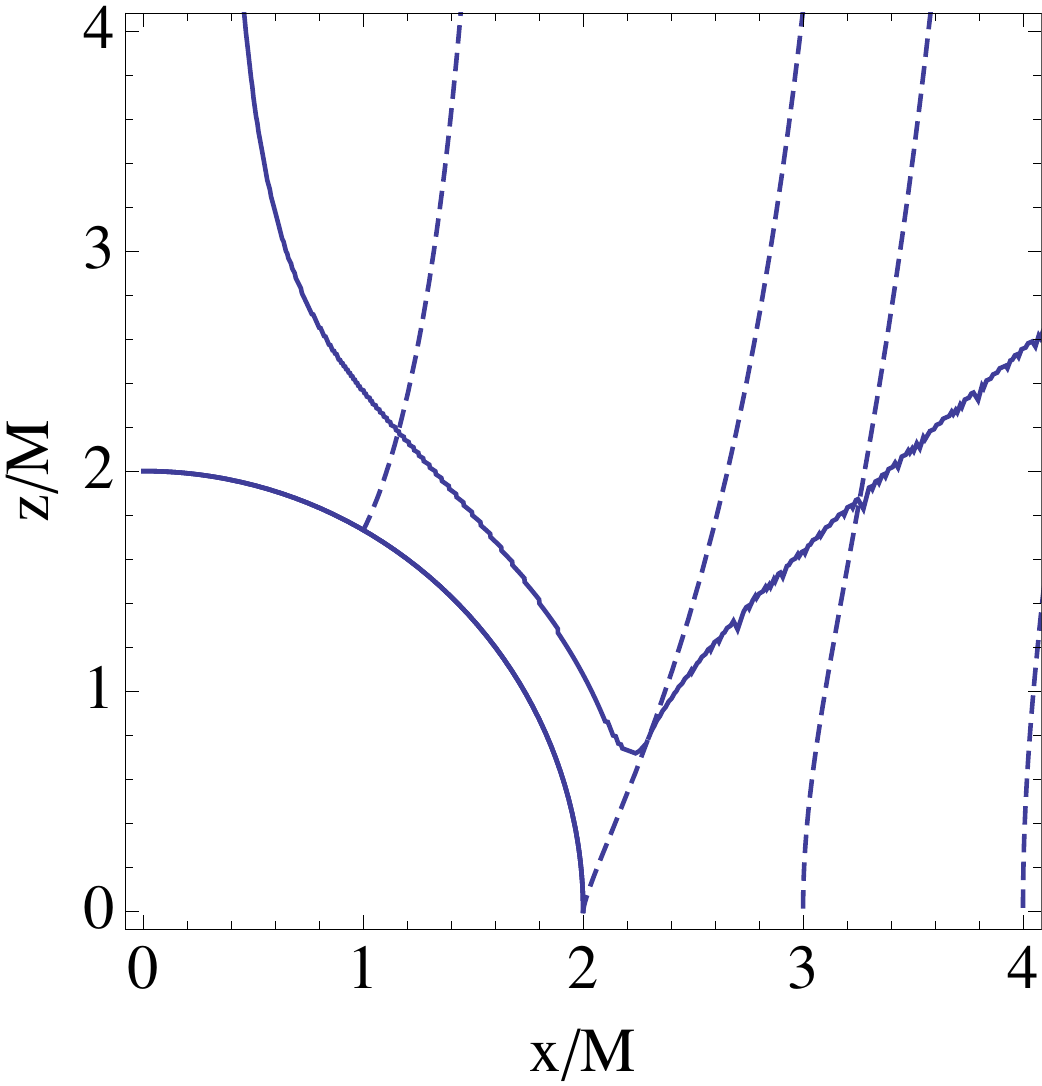}
   \caption{Structure of the equipotential surface in the $x-z$ plane (solid line). It is calculated by requiring that the potential drop along a given field
   lines equals some fiducial potential $\int E_\parallel ds= e \Delta \Phi_{\rm crit}$. Dashed lines are \Bf\ lines. 
   }
 \label{PFF}
 \end{figure}

    For a given parallel \Ef, Eq. (\ref{inv}) and a given shape of field lines, Eq. (\ref{shape}), we can calculated the total potential drop along a field line
    $\Delta \Phi = \int E_\parallel ds$, where $ds$ is taken along a given field line.   For magnetic field lines that intersect the \BH\ horizon (those that have $r_{\perp,0}  <e^2 M /2$) the lower limit of integration is on the horizon, while for those field lines than miss the hole the 
    lower limit of integration is at magnetic equator. After a particle gains sufficient energy $ e \Delta \Phi_{\rm crit}$ to break the vacuum via radiative effects, a dense flow of the secondary particles will screen the resulting parallel \Ef. We can calculate the shape of this so called pair formation front  (equipotential surface) by requiring $\int E_\parallel ds= e \Delta \Phi_{\rm crit}$, see Fig. (\ref{PFF}). 
    
Near the central field line the pair formation front is located at large distances from the \BH\, since close the axis the \Ef\  is small, 
 $E_\parallel \approx 2 E_0 \cos \phi { M r_\perp \over r^2}$, where $r_\perp$ is the  distance from the axis.
 To find the shape of a pair formation
  at large cylindrical distances,  we note that \Bf\ are nearly straight, while the parallel \Ef\ bcomes $E_\parallel \approx \sin 2 \theta \cos \phi ( M E_0/r)$. A fixed potential drop is then achieved on surfaces satisfying
  \be
  \sin \theta_{PPF} = {  e \Delta \Phi_{\rm crit} \over 2 M E_0 \cos \phi}.
  \ee
Formally, the pari formation front extends to infinity.

\subsection{Large scale jets: quadruple current flow}
    
\begin{figure}[h!]
   \includegraphics[width=0.99\linewidth]{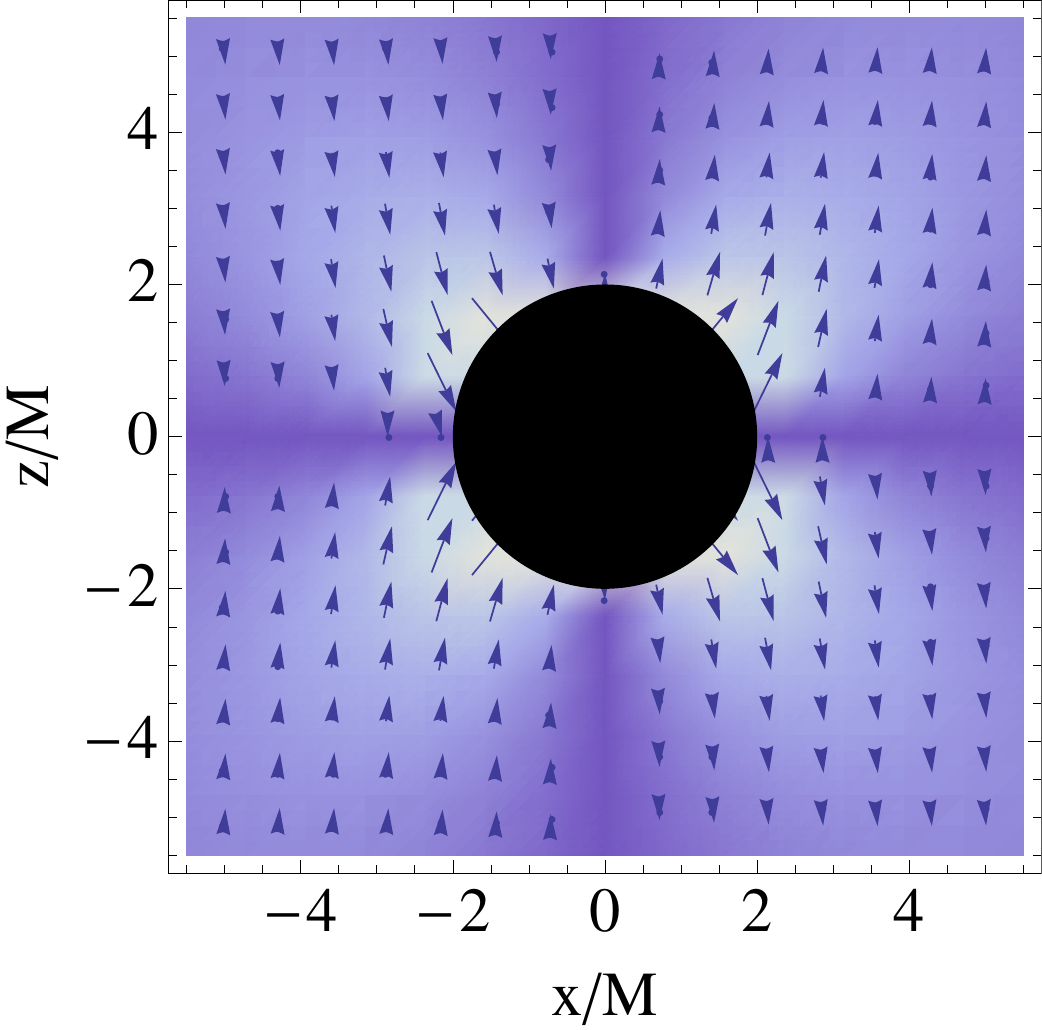}
   \caption{ Value and direction of parallel \Ef\ $E_\parallel$ in the plane $y=0$. Shadowing intensity is proportional to  the absolute value of the \Ef.
The \Ef\ has opposite direction in the four  quadrants;  as a result two counter-aligned current flows in each $\pm z$  direction will be generated.
   }
 \label{Eparaz2}
 \end{figure}

Accepting the paradigm  that the plasma will be streaming along \Bf\ lines nearly with the speed of light, the charge density  (\ref{rhoind}) will generate a current flow in the \BH\ \ms.
At large distances the sign of parallel \Ef\  and of the charge  density depends on the quantity $ \sin   \theta \cos \phi $, Eqns. (\ref{inv},\ref{rhoind}). Thus,  two counter-aligned  currents propagating along $z$ axis will be generated 
(four total separate current), separated the  $y-z$  plane, Fig. \ref{Eparaz2}.

Each quadrant of space carries a total current
\be
I \approx    r ^2 \int_{-\pi/2}^ {\pi/2} d\phi \int_0 ^{\pi/2} \sin \theta d\theta   \rho=  {\pi \over 4}   M^2   \rho_0 = { G M \beta_0 B_0 \over 16 }
\label{II}
\ee
This current will produce toroidal \Bf\ (for cylindrical radii $M \ll r_\perp \ll z$)
\be
B_\phi \approx \beta_0 {M r_\perp^2 \over 3 \pi z^3} \cos \phi B_0 \ll B_0
\label{Bphi}
\ee
At a given cross-section $z=constant$, the 
toroidal \Bf\ will roughly correspond to two equal counter-aligned current flows. 

The regions of space with two counter-aligned currents propagating along $\pm z$ direction (corresponding to semi-spaces $x> 0$ and $x<0$)  will interact with each other. Since the currents are due to charge-separated flow, the interaction is both electric and magnetic. Oppositely charged currents have charge density per unit length
$\lambda \approx M^2 \rho_0$. The total electric force per unit length will be $F_e \approx  \lambda ^2 /r = M^4 \rho_0^2 /r$. 
This electrostatic force will be nearly balanced (but not completely) by the magneto-static repulsive  force of two counter-aligned currents $I$, Eq. (\ref{II}). (The repulsive Lorentz force in this case will be smaller than the attractive Coulomb force by $1-v_b/c\ll 1$, where $v_b$ is the velocity of the particles from the primary beam.)

  \subsection{Outer gaps} 
\label{outer}

\begin{figure}[h!]
   \includegraphics[width=0.99\linewidth]{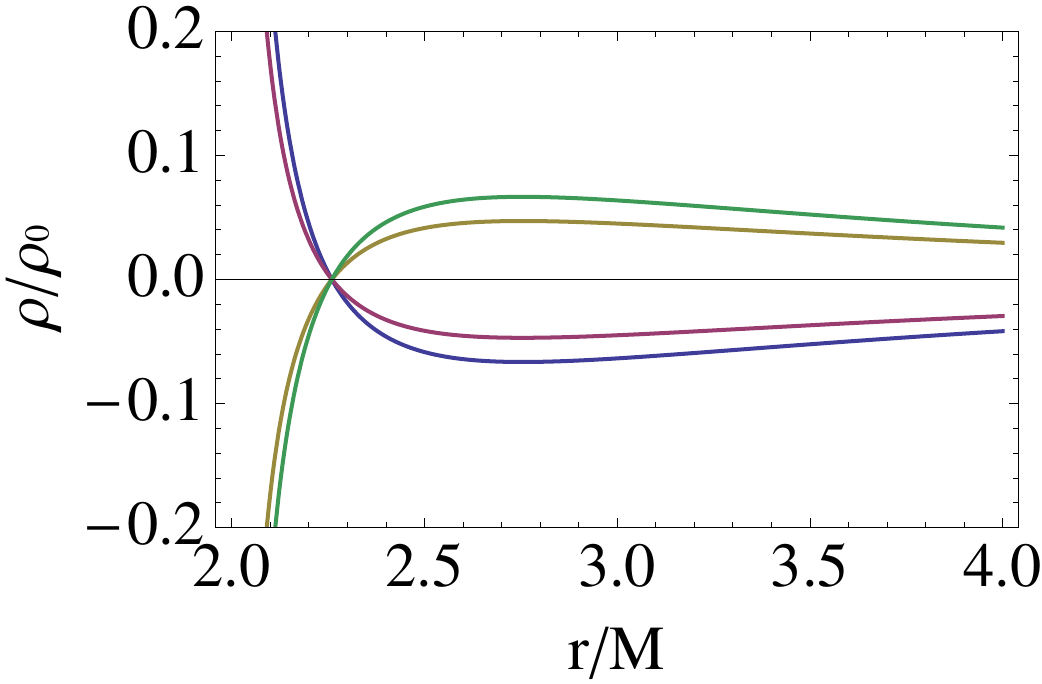}
   \caption{Charge density along \Bf\ field lines starting at $\theta =\pi/2$ at different azimuthal angles $\phi = -(3/4) \pi, - \pi,0, \pi/2 $ (top to bottom) at large $r$).
   The zero is at radii satisfying $\frac{4 M^2 e^{\frac{4 \sqrt{r-2 M}}{\sqrt{r}}} \left(\frac{M}{r}-\frac{3}{2}\right)}{\left(\sqrt{r-2
   M}+\sqrt{r}\right)^4}+1=0$, $r= 2.260 M$. 
   }
 \label{rhoalongB}
 \end{figure}

 On some field lines the charge density changes along a  field line, Fig. \ref{rhoalongB}.   By analogy with pulsar magnetospheres we will call the surfaces  $\rho_{\rm ind}=0$ the ``outer gaps''.
 The condition  $\rho_{\rm ind}=0$ is satisfied on surfaces given by 
 \be
 r_{og}= { 2 M \sin ^2 \theta \over 3 \sin ^2 \theta -2}
 \ee
 The outer gaps touch the horizon at the equator and extend within polar angles $ \arcsin \sqrt{2/3} <   \theta < \pi -  \arcsin \sqrt{2/3}$. 
  At large distances the outer gap becomes a conical surface with $   \cos 2 \theta = \pm 1/3$, see Fig. \ref{bigpicture}. 
In addition, some   \Bf\ lines have the sign of the induced charge density changing  two times along the field lines (this occurs along a narrow annular bunch of \Bf\ lines that cross the horizon close to the magnetic equator).

\section{\EM\ signals}

\subsection{Overall power}

What \EM\ signal can be expected from \Sc\ \BHs\ acting as a unipolar inductors?
In a magnetically dominated medium with the  effective impedance of free space, $\approx 4 \pi/c$,   the currents  (\ref{II})  will result in energy loss
\be
L _{EM, u} \approx M^2  E_0^2 = M^2 B_0^2 \beta_0^2
\label{LEM0}
\ee
The power (\ref{LEM0}) approximately equals $ B_\phi ^2 r^2 c $ with toroidal \Bf\ given by the estimate (\ref{Bphi}). Equivalently, it 
 arises due to initial \Ef\ and the toroidal \Bf\ (\ref{Bphi}) induced by the currents.

In particular, for a binary BH before merger, we can estimate the \Ef\ as 
\be
E_0 \sim B_0 \sqrt{ M/R}
\ee
where $R$ is the radius of the orbit. The resulting power is then
\be
L _{EM, u}= {M^3 B_0^2 \over R} \equiv {(G M)^3 B_0^2 \over c^5 R} 
\label{LEM2}
\ee
where we reinstated the Newton's constant and the speed of light. 
This is an estimate of the power lost by a \BH\ moving through constant \Bf\ via unipolar inductor mechanism. It turns out to be of the  same  order as 
$L_{EM,F}$, Eq. (\ref{L}),  the power dissipated by the rotation of space with the \BHs\ orbit via the 
Faraday disk mechanism 
\cite{2010arXiv1010.6254L}. Still, we stress that these powers comes from related but  physically separate mechanisms:  a Faradey-type \EM\ outflow  considered in \cite{2010arXiv1010.6254L}  and the linearly moving unipolar inductor discussed here (see also \cite{2010arXiv1012.2872M}). The power (\ref{LEM2}) is typically much smaller than the power carried away by gravitational waves at the later stages of the merger \cite{2010arXiv1010.6254L}.
 
\subsection{Estimates of  the \Bf}

 The general expression for the Poynting power (\ref{LEM2}) depends on the strength of the \Bf, which in case of merging astrophysical \BHs\   must be supported by the accretion disk and thus depend both on the microphysics of the disk dynamo \cite{BalbusHawley} and the dynamical evolution of the binary \BH-accretion disk system \cite{2005ApJ...622L..93M,2010ApJ...709..774K}.

Since the Poynting power (\ref{LEM2}) is the same as in the case of Faradey-type \EM\ outflow driven by the rotation of space-time, we can use various estimates of \Bf\  and the merger dynamics discussed in Ref.  \cite{2010arXiv1010.6254L}. Here we briefly restate the points. 

Two stages of disk dynamics may be identified. At early times the loss of energy via  gravitational radiation is slow enough, so that due to  viscous diffusion  the inner edge of the disk will be located close to the orbital radius, $R_{\rm d} \approx  R$. At later times, for $R_d \leq \xi_d G M /c^2, \, \xi _d \approx40- 100$  \cite{2005ApJ...622L..93M,2009ApJ...700..859O}, the binary will decouple from the disk, undergoing a   merger,  while the inner edge of the disk remains fixed at $R_d$. 

There are several estimates of \Bf, which generally give similar values  \cite{2010arXiv1010.6254L}. For example, 
 \Bf\ can be estimated assuming that   a fraction $ \eta_E$ of the Eddington luminosity is  carried by \Bf,   $B^2 \approx \eta_E L_{\rm Edd}/( c R_d^2)$, (where $R_d \sim R$ before decoupling and $R \sim \xi_d R_G$ after  decoupling)
\be
B =  \eta_{E}^{1/2} { (GM)^{1/2}  \sqrt{m_p} \over   \sqrt{ \sigma_T} R _d }  \approx    3 \times 10^4 \, {\rm G}\, m_6 ^{-1/2} \eta_{E,-1}^{1/2} \,  \left( {R_d \over R_G} \right)^{-1}
\ee
Thus, after decoupling \Bf\ remains nearly constant within the orbit, 
\be
 B_d \sim 300\, {\rm G} \, m_6 ^{-1/2} \eta_{E,-1}^{1/2} \, \xi_{d,2}^{-1}
 \label{Bd}
 \ee

The total power  (\ref{LEM2})  in this case then becomes
\be
{L_{EM}  \over L_{Edd}} =  \eta_{E}  {(G M)^3  \over c^6 R R_d^2 }  \leq \eta_{E} \,  \xi_d ^{-2} 
\label{30}
\ee
The peak \EM\ power is typically a fairly small fraction of the Eddington luminosity $L_{EM}  \approx 10^{-5} - 10^{-3} \,  L_{Edd}$ and can realistically be observed at cosmological distances only for merger of very massive \BHs, $M \geq 10^8M_\odot$.

Finally, comparing the power of the  \Sc\ \BH\ as unipolar inductor, Eq. (\ref{L}) with the Blandford-Znajek power in a given \Bf, $L_{BZ} \sim a^2 B^2 M^2$ ($a$ is the \BH\ spin parameter), 
 gives
 \be
 {L_{EM} \over L_{BZ}} \approx {M \over a^2 R}
 \ee
 Thus, for fast rotating {\BH}s, $a \sim 1$,  the power of the  unipolar inductor is subdominant for $R \gg M$.  (In addition, since during the merger the inner edge of the disk is at fairly large radii, $\geq 40-100 R_G$, it is expected that the \Bf\ on the \BH\ is smaller in the case of a merger).

\subsection{Expected emission}

Eqns (\ref{LEM2}-\ref{30})  provide  estimates of the  total  \EM\  power  produced by  a \BH\ in a form of a Poynting flux due to unipolar induction mechanism. A fraction of this power will be dissipated and converted into the observed radiation. Next, we discuss possible emission signatures. By analogy with pulsars, we expect several  types of \EM\ signal from \BHs\ as unipolar inductors,  approximately corresponding to magnetospheric and plerionic emission. In  addition, in case of \BHs\ we expect emission from  regions  with $E>B$.

 \subsubsection{Magnetospheric-type emission}

As we discussed above, the \mss\ of \BHs\  moving in \Bf\ resemble in many respects the pulsar \mss. Thus we may expect somewhat similar radiative signatures, though the details of the emission mechanisms proposed below must naturally be calculated independently for \BH\  \mss.
Rotationally-powered pulsars produce, generally speaking, two types of radiation: coherent radio emission and high energy $X$-ray through $\gamma$-ray emission. 

As we discussed in \S \ref{outer}, the 
 sign of the induced charge density changes as a  function of distance from the hole along a set of file lines, Figs. \ref{bigpicture},\ref{rhoalongB}.  One might expect that this will lead to effects qualitatively similar to the ones occurring in the so-called outer gaps in pulsar {\ms}s, where the sign of the Goldreich-Julian density changes  \cite{1986ApJ...300..500C}. In particular, recent \Fermi\ observations of pulsars show that high  energy $\gamma$-ray emission is generated at the outer gaps \cite{2010ApJS..187..460A}. Similarly, we can expect that merging \BHs\ will produce high energy emission, which will be preferentially beamed along the normal to the orbital plane.  One might expect that, similar to pulsars, the emitted high energy power may reach tens of percent of the total \EM\ power, Eq.  (\ref{LEM2}).

The radiation physics of pulsar outer gaps is not well understood at the moment \cite{2003ApJ...591..334H}.  A particularly important difference between the  pulsar and \BH\ \mss\ considered here is that in case of pulsars the curvature radiation (in addition to  inverse Compton processes) is an important ingredient. Magnetic fields in the magnetospheres of \BHs\ will generally have drastically different - larger - radii of curvature, that would make curvature radiation unimportant (this cane be easily demonstrated following the estimates of the inverse curvature emission given below). 

Another, in addition to curvature emission,  important ingredient in the dynamics of the pulsar outer gaps is the inverse Compton (IC) emission. It will provide  the dominant radiative friction effect to the particles accelerated in the \BH\ \mss. Let us next estimate its properties. First, balancing the radiation friction force $ \approx \gamma^2  e^4 U_{\rm ph} /( m_e^2 c^3)$, where $\gamma$ is the Lorentz factor of a particle and  $U_{\rm ph} $ is the radiation field energy density, with the acceleration rate by the induced \Ef, $e E_0 c$, the terminal Lorentz factor becomes
\be
\gamma = { m_e c^2 \sqrt{ E_0  \over e^3 U_{\rm ph}}}
\ee
Estimating \Ef\ as
$E= \beta_0 B_0 = \sqrt{R_G /R_{\rm orb} } B_0$ and relating the  radiation field energy density to Eddington luminosity,
$U_{\rm ph}= L_{Edd} / ( 4 \pi R_{\rm orb} ^2 c)$, we find
\be
\gamma = \left({  GM B m_e^2 \sigma _T \over e^3 m_p}  \right)^{1/2} \xi ^{3/4} = 6\times 10^6\ , B_4 ^{1/2} m_6 ^{1/2} \xi ^{3/4}
\ee
where, we remind,  $R_{\rm orb} = \xi r_G$. 

We can then estimate the IC power produce by the leptons with typical density  given by Eq. (\ref{rho0}) within a typical volume $\approx R_G^3$,
\be 
L_{IC}\approx  {\beta_0 \over 2 \pi \sqrt{\xi} }  {B^2 R_G^2 c}  =10^{40} \, {erg s ^{-1}} B_4^2 m_6^2 \xi^{-1}
\label{LIC}
\ee
The typical frequency of IC emission is given by up-scattering of thermal photons of energy emitted at the inner edge of the accretions disk with energy
\be
k_B T \approx \left( {G M \dot{M} \over 8 \pi \sigma_{SB} R^3} \right)^{1/4}
\ee
where $\sigma_{SB} $ is Stefan-Boltzmann constant, $k_B$ is  Boltzmann  constant  and $ \dot{M} $ is accretion rate. 
Scaling accretion rate $\dot{M}$ with Eddington luminosity, $\eta_M  \dot{M} c^2 = L_{Edd}$, where $\eta_M \sim 0.1$ is the accretion efficiency, we find
\be
T =  \left({ m_p c^5  \over  2 \eta_M   \sigma_T R^3} \right)^{1/4}\approx 120\, {\rm eV} \, m_6 ^{-1/4} \eta_M  ^{-1/4}  \xi ^{-3/4}
\ee
The corresponding IC photon energy is
\be
\epsilon \approx \gamma^2 T =\frac{B c^{5/4} G^{3/4} {k_B} M^{3/4} {m_e}^2 \xi ^{3/4} {\sigma _T}^{3/4}}{\sqrt[4]{2} e^3 {m_p}^{3/4}
   \sqrt[4]{{\eta _M}} \sqrt[4]{{\sigma _{SB}}}} = 800\, {\rm GeV} B_4 m_6^{3/4} \xi ^{3/4} \eta_{-1} ^{-1/4}
   \label{epsilon}
\ee
Luminosities (\ref{LIC}) at energies   (\ref{epsilon}) can be detected in reasonable time (\eg\ 100 hours of observations)  by atmospheric Cherenkov telescopes like HESS and VERITAS only within the few Mpc.

 Similarly to the case of pulsars,  merging BHs may also produce high brightness
  coherent radio emission. The parallel \Ef\ (\ref{Epara}) will produce 
 a primary beam of density (\ref{rho0}) moving with highly relativistic velocity. Pair production by the particles from primary beam will generate secondary plasma with much higher densities. Such momentum distribution of particles may result in  various types of plasma instabilities and generation of coherent high brightness radio emission \citep[\eg][]{lbm99,1999ApJ...521..351M}. 

 \subsubsection{Plerionic-type emission}

Most of the  power (\ref{LEM2}) will leave the \BH\ region in a form of relativistic highly magnetized wind. Even though the  instantaneous  power (\ref{LEM2})  is typically much smaller than the Eddington power corresponding to masses $M$, Eq. (\ref{30}), 
 the total released energy can be substantial  as we demonstrate below. 
% We can estimate the total energy released by a merging \BHs\ via unipolar inductor mechanism. 
Most of the energy is  released before decoupling. Indeed, after the decoupling
\Bf\ is constant within the orbit, Eq. (\ref{Bd}), so that  the power (\ref{LEM2}) integrated over the merging orbit with
$ R \approx \left( R_d^4 - {G^3  M^3 t \over c^5} \right)^{1/4}$  after decoupling gives
\be
E_{\rm d} =\eta_E  \int  {(G M)^3 B_d^2 \over c^5 R(t)} dt \approx  \eta_E   L_{\rm Edd} {R_d \over c} \approx B_d^2 R_d^3
\ee
 Thus, after decoupling  the \BHs\ spiraling in a constant \Bf\ dissipate  approximately the \Bf\ energy within the volume of the  decoupling radius. 

On the other hand the total energy released, before decoupling is 
\be
E_{\rm tot} \approx \eta_E { L_{\rm Edd} R_0 \over c} =  \eta_E  { G M m_p R_0 \over \sigma_T} 
\ee
where $R_0$ is the initial radius where the model becomes applicable. 
Thus, most of the energy is released at large separations of the \BHs. 

As an estimate of the initial radius $R_0$  where the model is applicable, one can compare the pressure created by the outflow, $\approx L_{EM} / (R_0^2 c)$ with the thermal energy density in the surrounding medium,  $\approx n_{\rm ex} m_p c_s^2$ ($ n_{\rm ex}$ is number density and $c_s$ is the speed of sound in the surrounding medium).
This gives
\ba && 
R_0 \approx \left( \eta_E { ( G  M)^4 \over c^6 c_s^2 n \sigma_T} \right)^{1/5}= 10^{-3} \, {\rm pc}\, m_6^{4/5} c_{s,7} ^{-2/5} n^{-1/5} \, \eta_{E,-1}^{1/5} 
\nn &&
E_{\rm tot} \approx \left( {  (GM)^{9} m_p^5 \over c^6 c_s^2 n \sigma_T^6}  \right)^{1/5} = 10^{47} \, {\rm erg} \, m_6^{9/5}  c_{s,6} ^{-2/5} \, \eta_{E,-1}^{6/5} 
n^{-1/5}
\label{R0}
\ea
This is a substantial amount of energy, especially for mergers of more massive \BHs.  It can in principal be observed both via direct emission and by appearance of dynamical morphological feature resembling a wind-blown nebular  in  the central parts of galaxies.

 \subsubsection{Emission from  from  regions  with $E>B$}

 Violation of the condition $B >E$  implies large electric field in plasma that cannot be reduced by the relative motion of charges
(recall that the drift velocity is independent of the sign of charge). Mass loading (inertia)  or dissipation will reduce \Ef\ to $E< B$. It is commonly assumed (mostly for the purposes of numerical simulations \cite{Spitkovsky06,2010PhRvD..82d4045P,2010Sci...329..927P}) that 
 the regions where $E>B$ will be strongly dissipative due to resistivity.  

We can estimate the volume where $E> B $ as $\approx R_G^3 \beta_0$.  Dissipation of the \EM\ fields inside the regions $E>B$ will lead to energy flux into those regions $L_{diss}  \approx R_G^2 E_0 B_0$ (this estimate neglects the fact that not all the surface of the \BH\ is covered by the regions with $E>B$, see Fig. \ref{EeqB}).  Thus, qualitatively, there will be local dissipation of the magnetic energy flux through the \Sc\ circle. For merging \BHs\ the resulting power
\be
L_{diss} \approx R_G^2 \beta_0 B_0^2= B_0^2  { ( G M) ^{5/2} \over c^4 R^{1/2}}
\label{Ldiss}
\ee
For $R\gg R_G$ this power is somewhat larger than the Poynting power of both the unipolar inductor and the Faraday wheel mechanisms (\ref{L}), yet the estimates that went into Eq. (\ref{Ldiss}) are likely to be solid upper limits due to the neglect of the geometry of the $E>B$ regions, which, in fact,  does not cover most of the \BH\ surface for $\beta_0 \ll 1$. 
In addition, a fraction of this power will be swallowed by the \BH, while  the escaping part will be heavily redshifted and unlikely to be observed.

\section{Force-free \mss\ of \Sc\ \BHs}
\label{FF}
We have argued above that the parallel \Efs\ created by the \BH\ motion lead to vacuum break down, generation of dense plasma that in turn  screens the parallel component of the \Ef. If the matter energy-density is much smaller that the energy-density of the \Bf, the plasma 
behavior will be controlled by \Bf, while nearly massless charge carriers provide the currents demanded by the dynamics and ensure 
 $\E\cdot \B = 0$ condition. This is called the force-free limit  \cite{Gruzinov99}.  Using the 
 $3+1$ formulation of the General Relativity  \cite{ThornMembrane}, Eq. (\ref{maxw1}) ,  taking  the total time derivative of the constraint $\E \cdot \B =0$ and eliminating  $D_t \E$ and   $D_t \B$ using Maxwell equations, one arrives at the corresponding  Ohm's law in Kerr  metric
 \be
 {\bf j} = {\left( \B \cdot \nabla \times (\alpha \B) -\E \cdot \nabla \times (\alpha \E)  \right)  \B  + \alpha (\nabla \cdot \E) \E \times \B \over 4 \pi \alpha B^2}
\label{GS00}
\ee
Note that this expression does not contain function $\vec{\beta}$.   Eq. (\ref{GS00})  is a non-linear equation for the time-dependent structure of  \mss\ of Kerr \BHs.
In the stationary, $\nabla \times (\alpha \E)=0$,   axisymmetric case  Eq. (\ref{GS00})  and Maxwell's equations  (\ref{maxw1})  reduce to the Grad-Shafranov equation in \Sc\ metric \cite{Beskin} for the $A_\phi$ component of the vector potential. In our case,  the lack of a cyclic variable prevents the reduction of Eq. (\ref{GS00}) to a single equation.

 \section{Discussion}

The electro-dynamics of 
\Sc\ \BHs\ moving through constant \Bf\ resembles in many respect the pulsar \mss. The motion of a \BH\
 in vacuum both generates non-zero \Ef\ along the \Bf, and, in addition, produces regions where $E>B$. 
 Similarly to the  case of rotationally powered pulsars, the non-zero parallel \Ef\ will lead to vacuum breakdown, generation of pair plasma and production of large-scale electric currents. Magnetospheres of moving \BHs\ will have pair formation fronts and outer gaps, where the sign of the induced charge density changes.
There is an important difference between pulsars and \BH: in case of pulsars the parallel \Ef\ is produced by real surface charges \cite{GoldreichJulian}, while in case of \BHs\ the parallel \Ef\ is a pure vacuum effect,  resulting  from the  curvature of  the space-time.

In case of merging \BHs\ we expect two kinds of \EM\ outflows: those originating in the close vicinity of the \BHs\ due to unipolar induction mechanism and those originating within the orbit due to the rotation of space-time. Both types of jets have  approximately  the same power. There is a  clear thought experiment where the two effects (due to rotation of space-time and due to linear motion of a \BH) differ. Consider a linear motion of a \BH\ through magnetic field. The effect discussed in this paper will appear, while the one discussed in Ref.  \cite{2010arXiv1010.6254L} would not. On the other hand, consider a rotating massive ring: in this case the effect considered in Ref.  \cite{2010arXiv1010.6254L}  will appear, but one considered in this paper would not.

The power of a unipolar inductor  is taken from the energy of the  linear  motion of the BH. Thus, there is an effective  friction force exerted by the \Bf\ onto the \BH. Qualitatively, there are two ways to create a  Poynting flux: (i)  in vacuum due to changing \EM\ fields; (ii) in plasma by generating  a current-carrying plasma outlfow. In vacuum there is no energy loss by a \BH: even though the magnetic fields are disturbed by the  passage of a \BH,  there is no wave emission, as can be seen from the fact that in the frame of the \BH\ the \EM\ fields are stationary.

In the case of plasma, no time-dependence is necessary in order to produce a  Poynting flux:  presence of plasma allows one  to chose from all the possible reference frames a one special frame, where  the \Ef\ is zero. Only in that special frame there is no Poynting flux, all other frames will have Poynting flux. Mathematically, in case of plasma the stress-energy tensor is diagonalizable, while this is generally not true in vacuum. 
Overall, the system under consideration  is very similar to  pulsars, in vacuum aligned rotator does not spin-down,  but in reality parallel \Efs\ generate currents that carries  \EM\ energy, extracted from rotation.

% (associated with classical field configurations Ð sphalerons  ( Klinkhamer F R and Manton N S 1984 Phys. Rev. D 30 2212)

Any attempt to simulate numerically the magnetospheres of moving {\BH}s  using MHD-type (fluid)  codes will face a problem of non-zero $\E\cdot \B \neq 0$ and the regions with $E > B$. Both these conditions 
violate a commonly used fluid assumptions. Pair production resulting from  $\E\cdot \B \neq 0$ will eventually ensure that 
$\E\cdot \B \approx 0$ in the bulk of the plasma. Thus, 
one possible way to simulate the \BH\ \mss\  (again following the work on pulsar \mss) is  to assume that the regions where the  ideal approximation is violated are sufficiently small, so that in the bulk the condition $\E \cdot \B=0$ is satisfied.  In the highly magnetized limit the dynamics then reduces to the force-free limit \S \ref{FF}. 
But experience in modeling pulsar \mss\  \cite{Spitkovsky06} tells us that in the highly magnetized limit the system would evolve towards violations of the ideal condition 
$\E\cdot \B = 0$ by creating current sheets,  where the magnetic field reverses, and, in addition, would spontaneously create regions with $E>B$. Resolving current sheets, or finding a proper prescription for treating them was a major obstacle in numerical simulations of pulsar \mss. The appearance of regions with $E>B$ required introduction of artificial resistivity and demonstrating that the final outcome is largely independent of these {\it ad hoc} numerical procedures). 
This was the approach taken in Ref.  \cite{2010PhRvD..82d4045P,2010Sci...329..927P,2010arXiv1012.5661N} who performed a number of  force-free simulations 
of \BH\ \mss. Our results are generally in agreement with Refs \cite{2010PhRvD..82d4045P,2010Sci...329..927P,2010arXiv1012.5661N}.

 Finally,  we note that  in the standard model of particle physics the  non-zero   second  Poincare  electromagnetic  invariant leads to the appearance of  sources of topological  axial vector currents that  can lead to the  local violation of the baryon and lepton numbers through the triangle
anomaly (which is responsible, \eg for the two photon decay of $\pi^0$). The triangle anomaly
violates baryon number through a nonperturbative
effect  \cite{1976PhRvL..37....8T,Rubakov}.
In case of a \BH\ moving through  a constant \Bf,  there is  a non-zero divergence of the electromagnetic topological current $J_\nu$
\ba &&
J_\nu = A^\mu  ({^*} F_{\mu\nu})
\nn &&
J_0 = {\bf A} \cdot {\bf B}=0
\nn &&
J_i= {\bf E}  \times {\bf A} + { A_0 \over \alpha}  {\bf B}
\nn &&
 J_{\mu;\mu} = -{7\over 4} \sin 2 \theta \cos \phi B_0 E_0 { M \over r} = {7\over 4} {\bf E} \cdot  {\bf B}
 \label{JJ}
\ea
Note that the helicity, the time component of the  topological current, is zero; the anomaly appears due to the 3-divergence of the spacial components of $J_\nu$.
 We leave a more detailed investigation of the implications for the  standard model of particle physics to a future work. 

I would like to thank Sergei Khlebnikov,  Sergei Komissarov,  Luis Lehner, Jonathan McKinney for discussions.

\bibliographystyle{apsrev}
% \bibliographystyle{plain}
%\bibliographystyle{apj}
%\bibliography{~/Home/PulsarRadio/PulsarBib}
\bibliography{/Users/maxim/Home/Research/BibTex}
%\bibliography{/Users/maximlyutikov/Home/Research/BibTex}
%\bibliography{~/Home/Research/HallNS}
%\bibliography{/Users/maximlyutikov/Home/Research/HallNS/HallNS}

\appendix

\section{Black hole \mss\ in Kerr-Schild coordinates}

Interpretation of results in General relativity is often a non-trivial exercise. The results of the main part of the article were obtained in \Sc\ coordinates.
\Sc\ coordinates have a singularity on the horizon, which often makes interpretation of the results problematic. To avoid the singularity the Kerr-Schild coordinates are often used instead. In this appendix we repeat the previous calculations in the Kerr-Schild coordinates and show that qualitatively the above-derived results generall hold.

The Kerr-Schild metric is given by 
 \be
 ds^2 = - \alpha^2 dt^2 + {4 M  \over r}  dt dr+ \left(  1 + {2 M \over r} \right)dr^2 + r^2 \left( d\theta^2 +  \sin ^2 \theta d\phi^2\right)
 \ee
The vacuum Maxwell equations (\ref{Maxwell}) then 
gives equation for $A_\phi$ the same  as  in \Sc\ coordinates, Eq.  (\ref{GS}). 
Thus, in Kerr-Schild coordinates the flux function corresponding to constant field at infinity is {A$_\phi $}$= (B_0/2) r^2 \sin ^2 \theta$. 

\begin{figure}[h!]
   \includegraphics[width=0.99\linewidth]{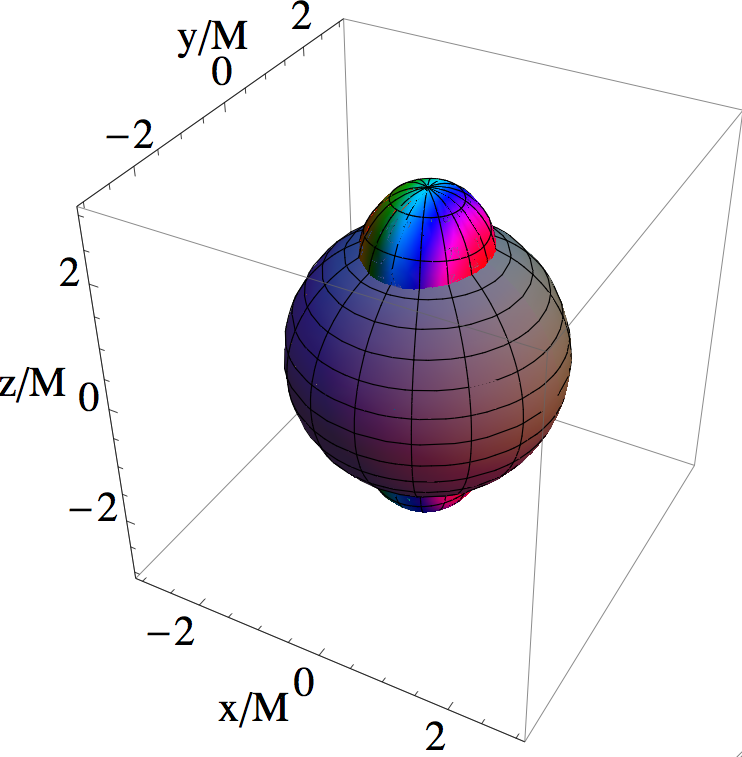}
   \caption{Shape of regions where $E=B$ in Kerr-Schild metric for $ \beta_0 =0.5$; compare with Fig. \ref{EeqB}.
   }
 \label{EeqBKS}
 \end{figure}

Using the covariant $\nabla$ operator with $
{\bf \hat{e}}_r = {1\over \sqrt{1+ 2M/r}} \partial_r
$ we can then find the  \EM\ fields and the charge density
\ba && 
\E = \sqrt{  1 + {2 M \over r} } \nabla A_0 = E_0 \left( { \cos \theta' }  {\bf e}_r  -   {1\over \sqrt{1+ 2M/r}}  \sin \theta'  {\bf e}_\theta \right)
\nn &&
\B =  {\nabla  \times A_\phi   {\bf e}_\phi} = B_0 \left(  { \cos \theta }   {\bf e}_r  -    {1\over \sqrt{1+ 2M/r}} \sin \theta  {\bf e}_\theta \right)
\nn &&
\rho _{\rm ind} =\frac{E_0 M \sin \theta  \cos \phi   \left(\left(\frac{M}{r}+\frac{3}{2}\right) \sin ^2(\theta
   )+\frac{M}{r}+1\right)}{\pi  \sqrt{\frac{2 M}{r}+1} \left(\frac{2 M \cos ^2\theta }{r}+1\right)^2}
\label{BBKS}
\ea
Qualitatively, the results in the Kerr-Schild metric look very similar to the ones in \Sc\ metric. 
The only noticeable difference is the location of the $E=B$ surface, which in the case of  the  Kerr-Schild metric  is given by 
\be
r=\frac{2 M \left(\sin ^2\theta -\beta_0 ^2 \left(\cos ^2\theta  \cos ^2\phi  +\sin ^2\phi  \right)\right)}{1-\beta_0
   ^2},
   \ee
   see Fig. (\ref{EeqBKS}). 

\end{document}